\documentclass[twocolumn,prc,superscriptaddress,amsmath,amssymb]{revtex4-1}%
\usepackage{epsfig,dsfont,amssymb,amsmath,amsthm,amsfonts,amsbsy,mathrsfs}
\usepackage{graphicx}
\usepackage{amsmath}
\usepackage{amssymb}%
\usepackage{booktabs}
\usepackage{multirow}
\usepackage{lineno}
\usepackage{dcolumn}
\usepackage{bm}
\usepackage{color}
\usepackage{float}

\usepackage[colorlinks,
           bookmarks=true,
           linkcolor=blue,
           urlcolor=blue,
            anchorcolor=black,
            citecolor=blue
            ]{hyperref}

\usepackage[all]{hypcap}

\begin{document}

\title{Understanding the Intermittency Signal in RHIC-STAR Data through Modeling}

\author{Jin Wu}
\affiliation{College of Physics and Electronic Information Engineering,\\
Guilin University of Technology, Guilin 541004, China}

\author{Zhiming Li}
\email{lizm@mail.ccnu.edu.cn}
\affiliation{Key Laboratory of Quark and Lepton Physics (MOE) and Institute of Particle Physics, \\
Central China Normal University, Wuhan 430079, China}

\author{Mingmei Xu}
\affiliation{Key Laboratory of Quark and Lepton Physics (MOE) and Institute of Particle Physics, \\
Central China Normal University, Wuhan 430079, China}

\author{Shuyun Yang}
\affiliation{School of Artificial Intelligence in Medicine,\\
Guilin Medical University, Guilin 541199, China}

\author{Ranran Guo}
\affiliation{Key Laboratory of Quark and Lepton Physics (MOE) and Institute of Particle Physics, \\
Central China Normal University, Wuhan 430079, China}

\author{Yuanfang Wu}
\email{wuyf@mail.ccnu.edu.cn}
\affiliation{Key Laboratory of Quark and Lepton Physics (MOE) and Institute of Particle Physics, \\
Central China Normal University, Wuhan 430079, China}


\begin{abstract}
Intermittency analysis provides a promising probe of scale-invariant density fluctuations near the QCD critical point. The intermittency measurements reported in the STAR BES-I data call for a quantitative assessment of the signal strength and a clearer physical understanding of its collision-energy dependence. In this work, we perform such a study for the STAR measurements using an improved hybrid UrQMD+CMC model, in which critical-like fluctuations are embedded into a realistic non-critical background through event-level, particle-level, and combined replacement schemes. By directly comparing the second-order factorial moment $\Delta F_{2}(M)$ between model calculations and experimental data on a point-by-point basis, we constrain the effective critical-like contribution compatible with the STAR measurements without relying on scaling exponents. The STAR data at $\sqrt{s_{\mathrm{NN}}}=7.7$--$27~\mathrm{GeV}$ used for model comparison can be consistently described only by small and nearly energy-independent effective critical-like fractions. These results indicate that the current BES-I intermittency signal is weak and exhibits little collision-energy dependence, thereby favoring only a limited critical-like contribution rather than a strong critical-point-induced enhancement localized near a specific collision energy.

\end{abstract}
	
\maketitle

\section{Introduction}

The quest to map the Quantum Chromodynamics (QCD) phase diagram of strongly interacting matter stands as a central challenge in heavy-ion physics~\cite{STARbesIIcumulant,QCDReport,NatureQGP,STARcumulant3GeV,NST,Luo:2022mtp}. In the search for the QCD critical point, intermittency analysis---a framework that examines the scaling behavior of scaled factorial moments (SFMs) over progressively finer divisions of phase space---has emerged as a sensitive probe of critical fluctuations~\cite{STARIntermittency,STARIntermittencyProce}. This method has been actively employed in the STAR experiment at RHIC~\cite{STARIntermittency,OurnPRpaper} and in the NA61/SHINE experiment at the CERN SPS~\cite{NA61SHINEArSC,ReynaOrtiz:2024hul,NA61SHINE:2023gez}. Recent results from the RHIC Beam Energy Scan Phase I (BES-I) have revealed, after background subtraction, a clear power‑law scaling of higher‑order SFMs with respect to the second‑order moment across all collision energies. In particular, in the most central (0–5\%) collisions, the extracted scaling exponent $\nu$ exhibits a non‑monotonic energy dependence, with a minimum around $\sqrt{s_\mathrm{NN}} = 27\ \text{GeV}$. Concurrently, no intermittency signal (i.e., no rise in the second-order SFM) was found for protons in Ar+Sc collisions at $\sqrt{s_\mathrm{NN}}$ =  5.1, 6.1, 7.6, 8.8, 11.9, and 16.8 GeV~\cite{NA61SHINEArSC}, while an increasing trend in the second-order correlator SFM was observed for negatively charged hadrons in central Xe+La collisions at $\sqrt{s_\mathrm{NN}}$ = 16.8 GeV~\cite{ReynaOrtiz:2024hul}. However, this complex experimental picture is still not fully understood, particularly with regard to the possible critical-like contribution to the intermittency signal at different collision energies and the question of which collision energy exhibits the largest signal in the current STAR BES-I data.

The ultra-relativistic quantum molecular dynamics (UrQMD) model \cite{MBUrQMD,SABassUrQMD,HPUrQMD,ELUrQMD,MBPRCUrQMD} provides a comprehensive description of the non-critical dynamics inherent to heavy-ion collisions, simulating the full collision process from initial transport to hadronic evolution. However, the UrQMD model alone does not exhibit any signature of intermittency in Au+Au collisions at RHIC energies~\cite{RefEfficiency,UrQMD+CMC,Intermittencysym}. Conversely, the critical Monte Carlo (CMC) model~\cite{AntoniouPRL,NGNPA2005,CMCPLB,LiOverview} is specifically designed to generate critical momentum-space fluctuations through Lévy random walks, yet it functions as a static momentum generator, lacking dynamical evolution. In prior studies~\cite{UrQMD+CMC,GuoEMs,Intermittencysym,WangPLB}, we developed an initial hybrid UrQMD+CMC framework by embedding CMC-generated critical fluctuations into the UrQMD hadronic background. This approach successfully reproduced the experimental scaling exponent when an intermittency signal of approximately 1--2\% was introduced.

In this work, we investigate the intermittency signal in RHIC-STAR data using an enhanced hybrid UrQMD+CMC model, which is designed to provide a more realistic description of possible critical fluctuations in heavy-ion collisions. Although our previous hybrid model represented an initial step toward embedding critical fluctuations into a non-critical hadronic background, it was restricted to a particle-level replacement scheme, in which only a small subset of particles in each UrQMD event was replaced with particles generated from the CMC model. Such a construction does not fully capture a physically motivated scenario: due to non-equilibrium dynamics~\cite{nonequilibrium-Nahrgang,nonequilibrium-Li}, finite-size and finite-time effects~\cite{Brofas:2026pup,PoberezhnyukPRC2020,AntoniouPRD}, only a subset of heavy-ion collision events, referred to here as critical events, may evolve sufficiently close to the critical region to develop appreciable critical fluctuations. In these critical events, most of the produced particles may, in principle, carry critical correlations. However, during the subsequent evolution from the chemical freeze-out stage to the final detected particle state, critical signals could be substantially diluted by non-critical dynamics and large background contributions, so that only a fraction of particles may retain detectable critical information. Consequently, a more realistic modeling framework, consistent with the expected physics of heavy-ion collisions, should incorporate both event-level and particle-level replacement schemes.

Furthermore, the methodological approach of our previous model-data comparison, which relied on the extraction of the scaling exponent, proved insufficient for drawing definitive conclusions~\cite{UrQMD+CMC}. The physical interpretation of the scaling exponent remains debated~\cite{PLBHypercontractivity,ReynaOrtiz:2025ldt}, and its theoretically expected critical value has not been firmly established, especially when the constraints of a realistic reduced transverse-momentum phase space and experimental detector acceptance are taken into account~\cite{STARIntermittency}. To circumvent this difficulty, the present study adopts a more direct and robust methodology: we perform a quantitative, point-by-point comparison of the scaled factorial moments between the STAR experimental data and the calculations from our improved UrQMD+CMC framework. In this sense, the present study is not simply aimed at reproducing the data, but rather at placing quantitative constraints on the possible contribution of critical fluctuations that remains consistent with the STAR measurements. By directly confronting the experimental measurements with the improved model calculations, and by incorporating the STAR Collaboration's key experimental observation regarding the QGP temperature at different collision energies~\cite{TemperatureQGP}, we provide new insights into the nature of the observed intermittency signal at RHIC intermediate energies.

This paper is organized as follows. In Sec.~II, we introduce the observables used in the intermittency analysis. In Sec.~III, we present the enhanced hybrid UrQMD+CMC framework, with particular emphasis on the implementation of the event-level and particle-level replacement schemes. In Sec.~IV, we present the apparent intermittency signal obtained from the UrQMD+CMC model for different event-level fractions. In Sec.~V, we perform a systematic comparison between the STAR measurements and the hybrid UrQMD+CMC calculations obtained with the particle-level and event-level replacement schemes, respectively, in order to constrain the effective critical-like contribution at different collision energies. In Sec.~VI, we further quantify the intermittency signal in the STAR data within the combined event- and particle-level replacement scheme. Finally, a summary of the main results and an outlook are given in Sec.~VII.

\section{Intermittency Analysis Method}
To probe scale-invariant density fluctuations in multiparticle production, Bialas and Peschanski ~\cite{Bialas:1985jb, Bialas:1988wc} introduced the concept of intermittency within the context of high-energy physics. This phenomenon is characterized by a power-law dependence of the Scaled Factorial Moments (SFMs) on decreasing scales when analyzing particle distributions in phase space. For a given order $q$, the SFM is denoted by $F_q(M)$ and defined in a $D$-dimensional phase space (e.g., momentum space) as~\cite{,Bialas:1985jb, Bialas:1988wc,AntoniouPRL,NA49EPJC,NGNPA2005,GLPRL}:
\begin{equation}
F_{q}(M)=\frac{\langle\frac{1}{M^{D}}\sum_{i=1}^{M^{D}}n_{i}(n_{i}-1)\cdots(n_{i}-q+1)\rangle}{\langle\frac{1}{M^{D}}\sum_{i=1}^{M^{D}}n_{i}\rangle^{q}},
\label{Eq:FM}
\end{equation}

\noindent where $M$ is the number of equal bins per dimension, resulting in a total of $M^{D}$ cells. The variable $n_{i}$ denotes the particle multiplicity within the $i$-th cell, and $\langle \cdots \rangle$ represents an average over an ensemble of collision events. 

A power-law behavior of $F_{q}(M)$ over a range of $M^{D}$ is regarded as a signature of intermittency, reflecting the scale-invariant structure of the dynamical density fluctuations, and is known as the $F_{q}(M)/M$ scaling~\cite{AntoniouPRL,AntoniouPRD,NA49EPJC,NA49PRC,NGNPA2001}:
\begin{equation}
F_{q}(M) \propto (M^{D})^{\phi_{q}}, \quad M\gg 1 .
\label{Eq:FqM}
\end{equation}

\noindent Here the exponent $\phi_{q}$ is called the intermittency index, whose magnitude reflects the strength of intermittency. A connection to the QCD critical point is provided by the 3D Ising universality class, which predicts specific critical values for the intermittency index: $\phi_{q}=\frac{5\times(q-1)}{6}$ for protons ($p$)~\cite{AntoniouPRL} and $\phi_{q}=\frac{2\times(q-1)}{3}$ for pions ($\pi$)~\cite{NGNPA2001,NGNPA2005}.

In heavy-ion collisions, however, the intermittency signal surviving in the final state is typically obscured by large background effects. To remove these background contributions at the level of factorial moments, a mixed-event technique is employed. Accordingly, a correlator moment $\Delta F_{q}(M)$ is defined as the difference between the SFMs obtained from the original data and those constructed from mixed events~\cite{NA49EPJC,NA49PRC,NA61universe,STARIntermittency,NGNPA2005}:
\begin{equation}
\Delta F_{q}(M) = F_{q}(M)^\text{data} - F_{q}(M)^\text{mix}.
\label{Eq:DeltaFq}
\end{equation}

In conventional intermittency analyses, after background subtraction, $\phi_q$ can be extracted from the scaling relation of $\Delta F_{q}(M)/M$, rather than from the original $F_{q}(M)/M$ scaling. It should be noted that another established scaling behavior~\cite{GLPRL, GLPRD, GLPRC, OchsPLB1988, Ochs1990ZPC, AMPTnu}, namely the power-law relation between the higher-order moment $\Delta F_{q}(M)$ and the second-order moment $\Delta F_{2}(M)$, expressed as $\Delta F_{q}(M) \propto \Delta F_{2}(M)^{\beta_{q}} $ for $M \gg 1$, and the associated scaling exponent $\nu$ derived from $\beta_{q} \propto (q-1)^{\nu}$, are not investigated in the present study. Instead, $\Delta F_{2}(M)$ is adopted as the primary observable, as it provides a more direct and robust measure of intermittency strength compared to the scaling exponent.

\section{The Hybrid UrQMD+CMC Model}

The ultra-relativistic quantum molecular dynamics model ~\cite{MBUrQMD,SABassUrQMD,HPUrQMD,ELUrQMD,MBPRCUrQMD} provides a microscopic description of heavy-ion collisions, simulating the non-equilibrium transport and hadronic evolution across a wide range of collision energies, from a few GeV up to the TeV scale reached at the CERN Large Hadron Collider. As UrQMD captures the essential non-critical dynamics while lacking an explicit implementation of the QCD phase transition and associated critical phenomena, it is a well-suited tool for simulating background phenomena. UrQMD calculations have shown no indication of $\Delta F_{q}(M)/M$ or $\Delta F_{q}(M)/\Delta F_{2}(M)$ scaling in Au+Au collisions at RHIC energies, as the obtained values of $\Delta F_{q}(M)$ are consistent with zero~\cite{UrQMD+CMC,Intermittencysym}.

To simulate the critical intermittency signal expected from self-similar (fractal) fluctuations in the vicinity of the QCD critical point, the critical Monte Carlo (CMC) model has been employed~\cite{AntoniouPRL,NGNPA2005,CMCPLB,UrQMD+CMC,LiOverview}. In the CMC approach, particle momenta are generated via L\'evy random walks, which naturally produce power-law correlations characteristic of critical dynamics. However, CMC is constructed as a static momentum-space generator and does not provide a full dynamical (time-evolving) description of the collision process. In our previous studies~\cite{UrQMD+CMC,Intermittencysym}, we therefore combined UrQMD and CMC into a hybrid UrQMD+CMC framework, which successfully reproduced the intermittency signal reported by the STAR experiment, including the rise of $\Delta F_{q}(M)$ with $M^{2}$ and the characteristic scaling of $\Delta F_{q}(M)/\Delta F_{2}(M)$~\cite{STARIntermittency}.

To constrain the possible contribution of critical fluctuations in experimental data, we employ a hybrid UrQMD+CMC framework, in which the non-critical background is described by UrQMD, while critical-like fluctuations are introduced in a controlled manner through the CMC model. The hybrid UrQMD+CMC event samples are constructed using a replacement procedure, in which selected particles and/or events from the UrQMD background sample are replaced by their CMC-generated counterparts. Accordingly, within the hybrid UrQMD+CMC framework, two key parameters are introduced:

\begin{enumerate}

\item Event-level replacement fraction ($\alpha_{e}$): Let the UrQMD background sample contain $N_{\mathrm{UrQMD}}$ events. In the event-level replacement scheme, a subset of these events, amounting to $N_{\mathrm{UrQMD}}\times\alpha_{e}$, is replaced by the same number of CMC events. Consequently, the resulting hybrid ensemble consists of two distinct classes of events: the remaining $(1-\alpha_{e})\times N_{\mathrm{UrQMD}}$ events are purely non-critical (UrQMD), while the replaced $\alpha_{e}\times N_{\mathrm{UrQMD}}$ events are fully critical (CMC).

Physically, $\alpha_{\mathrm{e}}$ characterizes the effective fraction of heavy-ion collisions whose dynamical trajectories pass sufficiently close to the critical region to develop appreciable critical fluctuations. Owing to non-equilibrium dynamics~\cite{nonequilibrium-Nahrgang,nonequilibrium-Li}, as well as finite-size and finite-time effects~\cite{Brofas:2026pup,PoberezhnyukPRC2020,AntoniouPRD}, the growth of critical fluctuations in realistic heavy-ion collisions is expected to be strongly suppressed. Consequently, observable critical fluctuations are unlikely to be distributed uniformly over the whole event ensemble, but are expected to be concentrated in a limited subset of collisions.

\item Particle-level replacement fraction ($\alpha_{p}$): The particle-level replacement scheme is characterized by the replacement ratio
\begin{equation}
\alpha_{p}=\frac{n_{\mathrm{CMC}}}{n_{\mathrm{UrQMD}}},
\end{equation}
where $n_{\mathrm{UrQMD}}$ denotes the multiplicity ($N_{ch}$) of a given UrQMD background event, and $n_{\mathrm{CMC}}$ is the number of particles selected from a CMC event and embedded into that UrQMD event. The resulting hybrid event therefore contains two components: a non-critical background of $(1-\alpha_{p}) \times n_{\mathrm{UrQMD}}$ particles from UrQMD and a critical component of $\alpha_{p} \times n_{\mathrm{UrQMD}}$ particles from CMC.

Physically, $\alpha_{\mathrm{p}}$ quantifies the effective fraction of particles in a critical event for which critical information survives throughout the entire collision evolution and is retained in the final-state particles reaching the detector, rather than being erased by substantial background effects and non-critical dynamical noise. Even in critical events that pass close to the critical region, the primordial critical fluctuations can be substantially attenuated by non-critical dynamics and background effects, including trivial statistical fluctuations, global conservation laws~\cite{BzdakPRC2013}, resonance decays and hadronic rescattering~\cite{ZhangPRC2020}, finite acceptance effects~\cite{LingPRC2016,BzdakPRC2017,STARPRCMoment}, as well as finite momentum resolution~\cite{SamantaJPG2021}.

\end{enumerate}

It should be emphasized that the two replacement fractions, $\alpha_{e}$ and $\alpha_{p}$, are not merely technical inputs of the hybrid model. Instead, they are introduced as effective phenomenological parameters that quantify, respectively, the event-level occurrence probability of critical events and the particle-level survival fraction of critical information within events. Through a direct comparison between the hybrid-model calculations and the STAR measurements, these parameters can be quantitatively constrained, thereby providing an estimate of the critical-like contribution compatible with the experimental data.

The detailed procedure for generating the hybrid UrQMD+CMC events is outlined as follows:
\begin{enumerate}
\item UrQMD background generation. Generate a sample of UrQMD events with the desired model settings. For each event, record the charged-particle multiplicity $N_{\mathrm{ch}}$ and the momenta of all final-state particles.

\item CMC critical-event construction. Configure the CMC model with the L\'evy parameter $\mu = 1/6$ and $p_{\mathrm{min}}/p_{\mathrm{max}} = 10^{-7}$. These choices correspond to a critical system in the 3D Ising universality class with fractal dimension $d_{F}=1/3$~\cite{AntoniouPRL,CMCPLB}. For each UrQMD event, randomly select one UrQMD particle with $p_{T}<0.5~\mathrm{GeV}/c$ as the initial point of a L\'evy random walk. The random walk then produces a set of particle momenta ($p_{x},p_{y}$) that constitutes a CMC event, constructed to match the UrQMD event in charged-particle multiplicity and to reproduce, to good approximation, its transverse-momentum spectrum.

\item Hybridization via replacement schemes. Construct the UrQMD+CMC sample by applying one of the following replacement schemes:
\begin{enumerate}
    \item Event-level replacement scheme only. Randomly select an event from the UrQMD sample, and then replace this entire event with its matched CMC event. Repeat until the number of replaced events reaches $\alpha_{e}\times N_{\mathrm{UrQMD}}$. The hybrid sample is then formed by the remaining UrQMD events together with the inserted CMC events.

    \item Particle-level replacement scheme only. For each UrQMD event, replace a fraction $\alpha_{p}$ of its particles with particles from the matched CMC event. In practice, we randomly sample one candidate particle from the UrQMD event and one from the corresponding CMC event, and accept the replacement only if their transverse momenta satisfy
\begin{equation}
\left|p_{T}^{\mathrm{CMC}}-p_{T}^{\mathrm{UrQMD}}\right| < 0.2~\mathrm{GeV}/c.
\end{equation}
    If the criterion is not satisfied, repeat the random sampling until an acceptable pair is found. After a successful replacement, remove the used CMC particle from the CMC pool to avoid reuse. The $p_{T}$-matching requirement is imposed to preserve the inclusive $p_{T}$ spectrum of the original UrQMD event in the hybrid sample. Continue replacements until the number of substituted particles reaches $\alpha_{p}\times N_{\mathrm{ch}}$ for that event, and apply the same procedure to all UrQMD events to obtain the full hybrid dataset.

    \item Combined event- and particle-level replacement scheme. In the combined replacement scheme, an $\alpha_{e}$ fraction of UrQMD events is first randomly selected from the background sample. For each selected event, the particle-level replacement procedure described above is then applied: a fraction $\alpha_{p}$ of UrQMD particles is replaced by particles from the corresponding matched CMC event. This procedure is repeated until the number of selected and modified events reaches $\alpha_{e}\,N_{\mathrm{UrQMD}}$, while all non-selected events remain as pure UrQMD background. The resulting hybrid ensemble thus contains two classes of events: a fraction $(1-\alpha_{e})$ of purely non-critical UrQMD events and a fraction $\alpha_{e}$ of hybrid events in which critical-like fluctuations are embedded at the particle level with strength $\alpha_{p}$. In this way, $\alpha_{e}$ controls the occurrence probability of critical events within the heavy-ion event ensemble, while $\alpha_{p}$ controls the fraction of particles within such events that retain critical information.

\end{enumerate}

\end{enumerate}

In this work, we employ UrQMD (version~3.4) in cascade mode to generate Au+Au collision events at RHIC energies. The UrQMD event statistics are chosen to be comparable to those used in the STAR intermittency measurements, amounting to $3.96\times10^{6}$, $8.80\times10^{6}$, $18.0\times10^{6}$, and $35.4\times10^{6}$ events at $\sqrt{s_{\mathrm{NN}}}=7.7$, $11.5$, $19.6$, and $27~\mathrm{GeV}$, respectively. For the construction of the UrQMD+CMC event sample, CMC events are generated with the same event statistics as the corresponding UrQMD samples. The analysis strategy closely follows that of the STAR intermittency measurements, including the centrality definition, kinematic selections, analysis acceptance, and fitting procedure~\cite{STARIntermittency,OurnPRpaper}. Charged hadrons (protons $p$, antiprotons $\bar{p}$, kaons $K^{\pm}$, and pions $\pi^{\pm}$) are selected within $|\eta|<0.5$. The transverse-momentum ranges are $0.2<p_{T}<1.6~\mathrm{GeV}/c$ for $K^{\pm}$ and $\pi^{\pm}$, and $0.4<p_{T}<2.0~\mathrm{GeV}/c$ for $p$ and $\bar{p}$. To suppress autocorrelations between the centrality estimator and the particles used in the intermittency analysis, centrality is determined from the uncorrected charged-particle multiplicity measured in $0.5<|\eta|<1.0$, i.e., outside the analysis window $|\eta|<0.5$. For the intermittency calculation, the transverse-momentum plane $(p_{x},p_{y})$ within $[-2.0<p_{x}<2.0~\mathrm{GeV}/c]\otimes[-2.0<p_{y}<2.0~\mathrm{GeV}/c]$ is partitioned into $M^{2}$ equal-sized cells, with $M$ varied from 1 to 100.

\begin{figure*}
     \centering
     \includegraphics[scale=0.85]{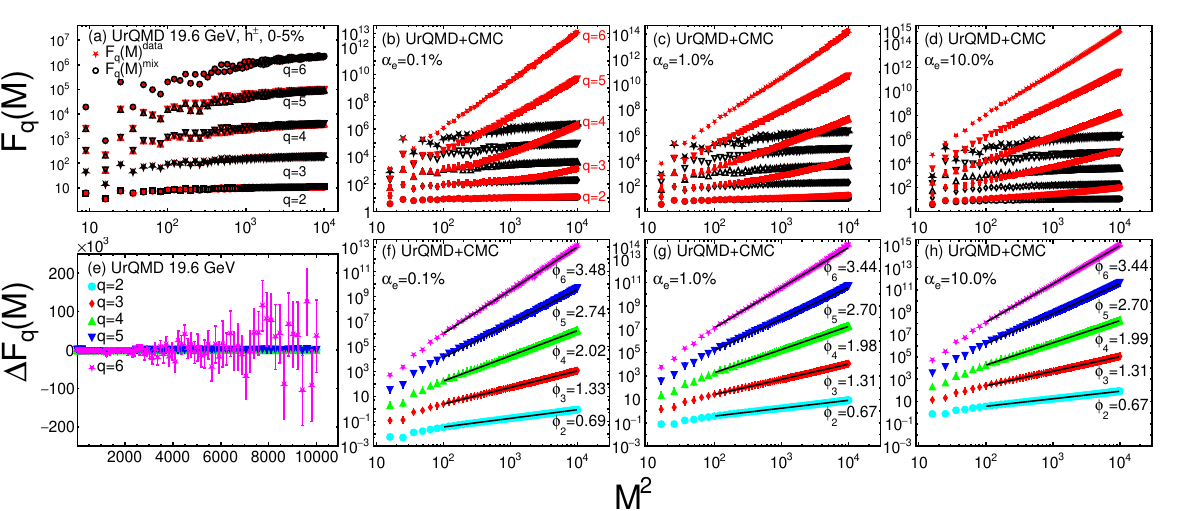}
      \caption{(a) Scaled factorial moments $F_{q}(M)$ ($q\leq 6$) for charged hadrons in the 0--5\% most central Au+Au collisions at $\sqrt{s_{\mathrm{NN}}}=19.6~\mathrm{GeV}$ generated with the UrQMD model, shown as functions of the number of phase-space cells ($M^{2}$) on double-logarithmic axes. The results for the original events ($F_{q}(M)^{\mathrm{data}}$) and the corresponding mixed events ($F_{q}(M)^{\mathrm{mix}}$) are displayed with red solid and black open markers, respectively. (b--d) $F_{q}(M)^{\mathrm{data}}$ and $F_{q}(M)^{\mathrm{mix}}$ as functions of $M^{2}$ for the hybrid UrQMD+CMC model with event-level critical-like fractions $\alpha_{\mathrm{e}}=0.1\%$, $1.0\%$, and $10.0\%$, respectively. (e) The correlator moment $\Delta F_{q}(M)$ versus $M^{2}$ at $\sqrt{s_{\mathrm{NN}}}=19.6~\mathrm{GeV}$ from the UrQMD model. (f--h) $\Delta F_{q}(M)$ versus $M^{2}$ for the hybrid UrQMD+CMC model with $\alpha_{\mathrm{e}}=0.1\%$, $1.0\%$, and $10.0\%$, respectively. The extracted $\phi_q$ values are quoted next to the corresponding fit lines.}
     \label{Fig1:UrQMD+CMCEvent_level}
\end{figure*}

\section{Intermittency Signatures from the UrQMD+CMC Model}

To reproduce the intermittency signal reported by the STAR experiment---in particular, the increase of $\Delta F_{q}(M)$ with $M^{2}$---we first construct hybrid UrQMD+CMC samples using the event-level replacement scheme. Figure~\ref{Fig1:UrQMD+CMCEvent_level}(a) shows $F_{q}(M)$ for charged hadrons from UrQMD events (red solid symbols) and the corresponding mixed events (black open symbols) as functions of $M^{2}$ for the 0--5\% most central Au+Au collisions at $\sqrt{s_{\mathrm{NN}}}=19.6~\mathrm{GeV}$. The corresponding correlator moments $\Delta F_{q}(M)$ ($q=2$--6), calculated according to Eq.~\eqref{Eq:DeltaFq}, are shown in Fig.~\ref{Fig1:UrQMD+CMCEvent_level}(e). For the pure UrQMD baseline, $F_{q}(M)^{\mathrm{data}}$ nearly overlaps with $F_{q}(M)^{\mathrm{mix}}$ over the full $M^{2}$ range, and $\Delta F_{q}(M)$ remains consistent with zero within statistical uncertainties. This behavior is expected, since UrQMD does not include density fluctuations associated with the QCD phase transition.

We then progressively embedded critical CMC events into the UrQMD background by varying the event-level critical-like fraction over $\alpha_{\mathrm{e}}=0.1\%$, $1.0\%$, and $10.0\%$. As shown in Fig.~\ref{Fig1:UrQMD+CMCEvent_level}(b)--(d), even for the smallest fraction $\alpha_{\mathrm{e}}=0.1\%$, $F_{q}^{\mathrm{data}}(M)$ becomes visibly larger than $F_{q}^{\mathrm{mix}}(M)$, with the separation increasing with moment order $q$. The correlator moments $\Delta F_{q}(M)$ ($q=2$--6), displayed in Fig.~\ref{Fig1:UrQMD+CMCEvent_level}(f)--(h), exhibit an approximately linear dependence on $M^{2}$ over the $M^{2}$ range for all three values of $\alpha_{\mathrm{e}}$, consistent with an intermittency-like power-law behavior. For $\alpha_{\mathrm{e}}=0.1\%$, a small deviation from the ideal $\Delta F_{q}(M)/M$ scaling is observed at the highest order ($q=6$); however, the second-order correlator moment $\Delta F_{2}(M)$ is largely unaffected, yielding an intermittency index $\phi_{2}=0.69\pm0.10$, which remains close to the critical expectation $\phi_{2}=2/3$.

As the signal fraction increases from $0.1\%$ to $1.0\%$ and further to $10.0\%$, the magnitudes of both $F_{q}(M)$ and $\Delta F_{q}(M)$ increase accordingly, with the ordinate ranges expanding by approximately one order of magnitude for each tenfold increase in the signal fraction. This monotonic dependence establishes a direct correspondence between the magnitude of $\Delta F_{q}(M)$ and the effective critical-like fractions, thereby enabling quantitative constraints to be extracted from experimental data. Importantly, this behavior also implies that, even when the intermittency index $\phi_{q}$ cannot be reliably extracted because the power-law scaling $\Delta F_{q}(M)\propto (M^{2})^{\phi_{q}}$ does not hold over the full $M^{2}$ range, the relative strength of intermittency in experimental data can still be qualitatively assessed from the magnitude of the measured $\Delta F_{q}(M)$.

\section{Constraining Critical-like Fractions with Individual Event- and Particle-Level Replacement Schemes}

\begin{figure*}
     \centering
     \includegraphics[scale=0.85]{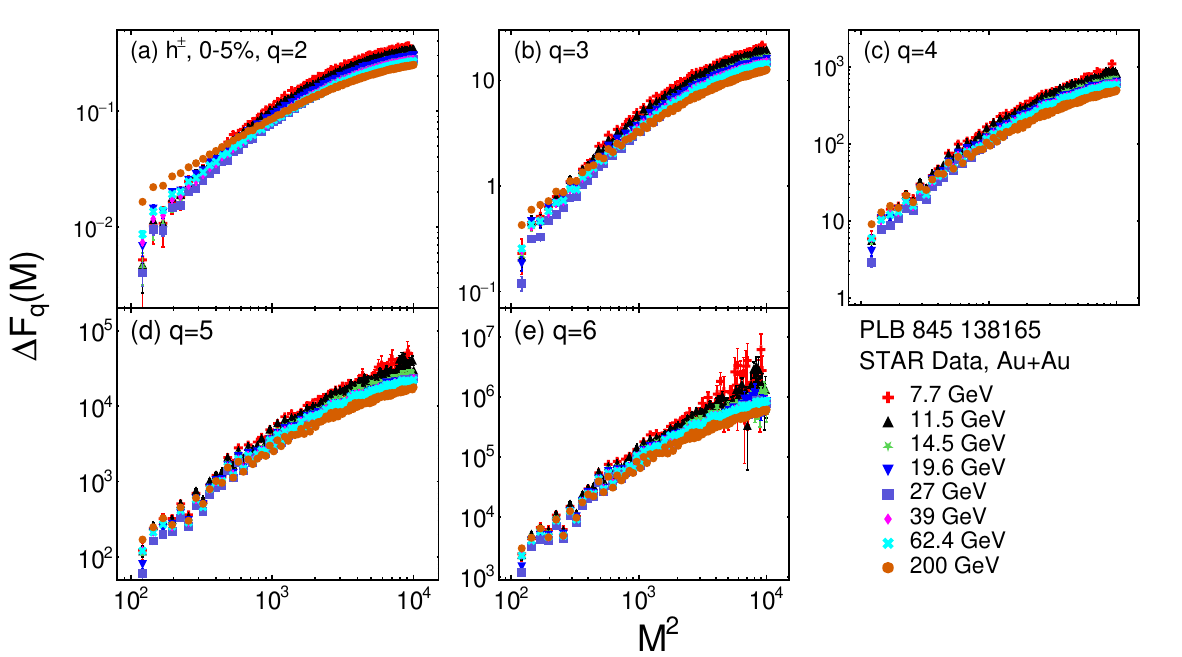}
      \caption{The correlator moments $\Delta F_{q}(M)$ as functions of $M^{2}$ for charged hadrons in the 0--5\% most central Au+Au collisions at $\sqrt{s_{\mathrm{NN}}}=7.7$, $11.5$, $14.5$, $19.6$, $27$, $39$, $62.4$, and $200~\mathrm{GeV}$ measured in the STAR BES-I data~\cite{STARIntermittency}. Panels (a)--(e) present the results for $q=2$--6, respectively. The data points shown here are taken from Ref.~\cite{STARIntermittency}.}
     \label{Fig2:STARFqVsEnergy}
\end{figure*}

We first replot the STAR BES-I data~\cite{STARIntermittency} for $\Delta F_{q}(M)$ at different collision energies, as shown in Fig.~\ref{Fig2:STARFqVsEnergy}. The figure presents $\Delta F_{q}(M)$ for fixed moment orders $q=2$--$6$ in panels~(a)--(e), respectively, and compares the results across the full STAR BES-I energy range, $\sqrt{s_{\mathrm{NN}}}=7.7$--$200~\mathrm{GeV}$. For each order $q$, the overall magnitude of $\Delta F_{q}(M)$ at large partition numbers, particularly for $M^{2}>1000$, shows a mild decreasing trend with increasing $\sqrt{s_{\mathrm{NN}}}$, with the largest values observed at $\sqrt{s_{\mathrm{NN}}}=7.7~\mathrm{GeV}$ and the smallest values at $\sqrt{s_{\mathrm{NN}}}=200~\mathrm{GeV}$. Nevertheless, the differences among the various collision energies remain relatively small, and the data points exhibit an approximate overlap over the measured $M^{2}$ interval. This limited energy variation suggests that $\Delta F_{q}(M)$ has only a weak dependence on collision energy, with its strength constrained within a relatively narrow range.

In our previous work~\cite{UrQMD+CMC}, the comparison between the hybrid UrQMD+CMC model and the STAR data was primarily based on the extracted scaling exponent. However, the theoretically expected critical value of this exponent has not yet been firmly established, and its connection to the QCD critical point remains debated~\cite{PLBHypercontractivity,ReynaOrtiz:2025ldt}. In the present work, instead of using the scaling exponent, we perform a direct and quantitative comparison of $\Delta F_{2}(M)$ between the UrQMD+CMC results and the STAR data. Compared with higher-order moments, $\Delta F_{2}(M)$ is less prone to being distorted by large non-critical background contributions when the intermittency signal is weak, and therefore provides a more stable observable for constraining small critical-like components.

\begin{figure*}
     \centering 
     \includegraphics[scale=0.90]{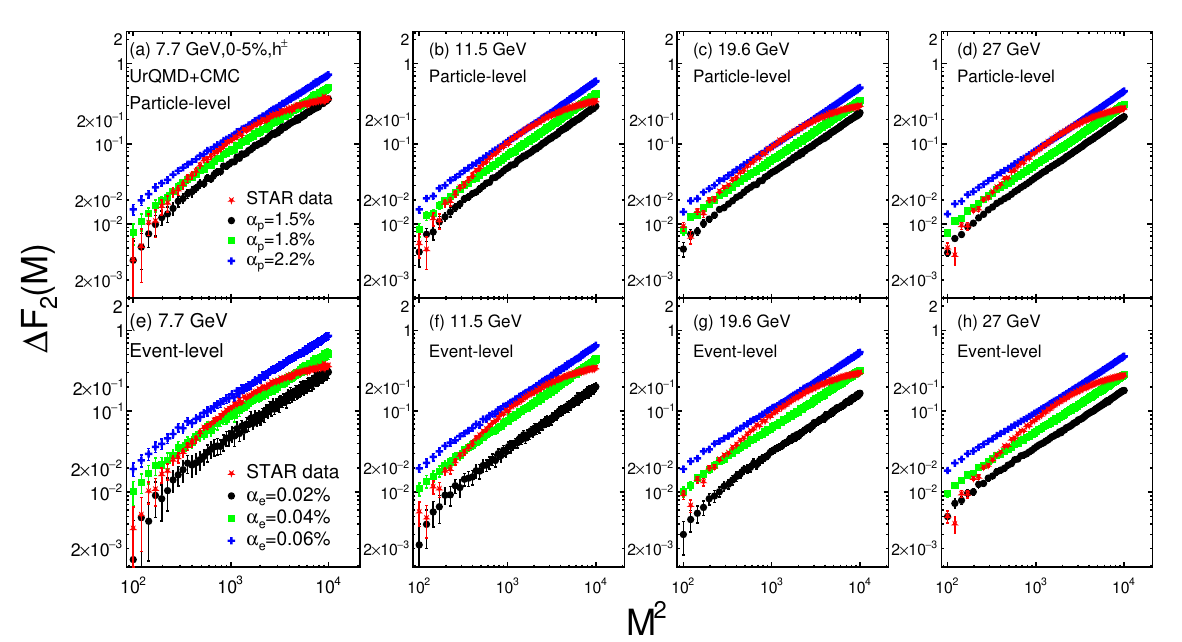}
      \caption{The upper panels, (a)--(d), show the comparison between the STAR data~\cite{STARIntermittency} and the hybrid UrQMD+CMC calculations obtained with the particle-level replacement scheme at $\sqrt{s_{\mathrm{NN}}}=7.7$, $11.5$, $19.6$, and $27~\mathrm{GeV}$, respectively, for $\alpha_{\mathrm{p}}=1.5\%$, $1.8\%$, and $2.2\%$. The lower panels, (e)--(h), present the corresponding comparisons for the event-level replacement scheme at the same collision energies, with $\alpha_{\mathrm{e}}=0.02\%$, $0.04\%$, and $0.06\%$.}
     \label{Fig3:ParticleLevel_and_EventLevel}
\end{figure*}

In the upper panels of Fig.~\ref{Fig3:ParticleLevel_and_EventLevel}, we compare the STAR data with the UrQMD+CMC results obtained using the particle-level replacement scheme at $\sqrt{s_{\mathrm{NN}}}=7.7$, $11.5$, $19.6$, and $27~\mathrm{GeV}$, while the corresponding comparisons for the event-level replacement scheme are shown in the lower panels. For the particle-level replacement scheme, the STAR data, shown as red stars, are systematically bracketed by the model calculations with $\alpha_{\mathrm{p}}=1.5\%$ and $2.2\%$, and exhibit the best overall consistency with the simulation using $\alpha_{\mathrm{p}}\simeq 1.8\%$ at the considered collision energies. For the event-level replacement scheme, the experimental data are similarly bounded by the model results with $\alpha_{\mathrm{e}}=0.02\%$ and $0.06\%$, with the best overall agreement obtained for $\alpha_{\mathrm{e}}\simeq 0.04\%$ at the considered collision energies. These comparisons therefore indicate that the STAR measurements restrict the effective particle-level and event-level critical-like fractions to narrow ranges, favoring $\alpha_{\mathrm{p}}\simeq 1.8\%$ and $\alpha_{\mathrm{e}}\simeq 0.04\%$, respectively. The extremely small value of $\alpha_{\mathrm{e}}$ further suggests that, within the event-level interpretation, only a tiny subset of heavy-ion collisions may develop a fully critical-like component whose critical intermittency signal survives the subsequent dynamical evolution and remains observable in the final state.

Despite the overall quantitative agreement in the magnitude of $\Delta F_{2}(M)$, a residual discrepancy remains between the hybrid UrQMD+CMC calculations and the STAR data in the shape of $\Delta F_{2}(M)$ for both the event-level and particle-level replacement schemes. In the present hybrid model, $\Delta F_{2}(M)$ exhibits an approximately linear dependence on $M^{2}$ over the full range considered, whereas the STAR data show a clear upward curvature at intermediate $M^{2}$, followed by a tendency toward saturation at larger $M^{2}$. This systematic difference in the shape of $\Delta F_{2}(M)$ indicates that the current implementation of the hybrid model may not fully capture the underlying structure of critical fluctuations, and may point to more complex dynamics, possibly associated with three-dimensional, self-affine critical behavior. In realistic heavy-ion collisions, critical fluctuations are expected to develop in the full three-dimensional momentum space and to exhibit a self-affine, rather than a strictly self-similar geometry~\cite{WuPRL,PLBNA22}. Under such circumstances, even if a power-law behavior holds in the full three-dimensional phase space, its projection onto the two-dimensional $(p_{x},p_{y})$ plane may generate a non-linear, curved dependence of $\Delta F_{2}(M)$ when analyzed within the conventional self-similar framework. Therefore, an improved hybrid model incorporating three-dimensional self-affine critical fluctuations would be desirable for a more realistic description of the measured $\Delta F_{2}(M)$ and for a more reliable characterization of the critical intermittency signal.

\section{Constraining Critical-like Fractions with the Combined Event- and Particle-Level Replacement Scheme}

\begin{figure*}
     \centering 
     \includegraphics[scale=0.66]{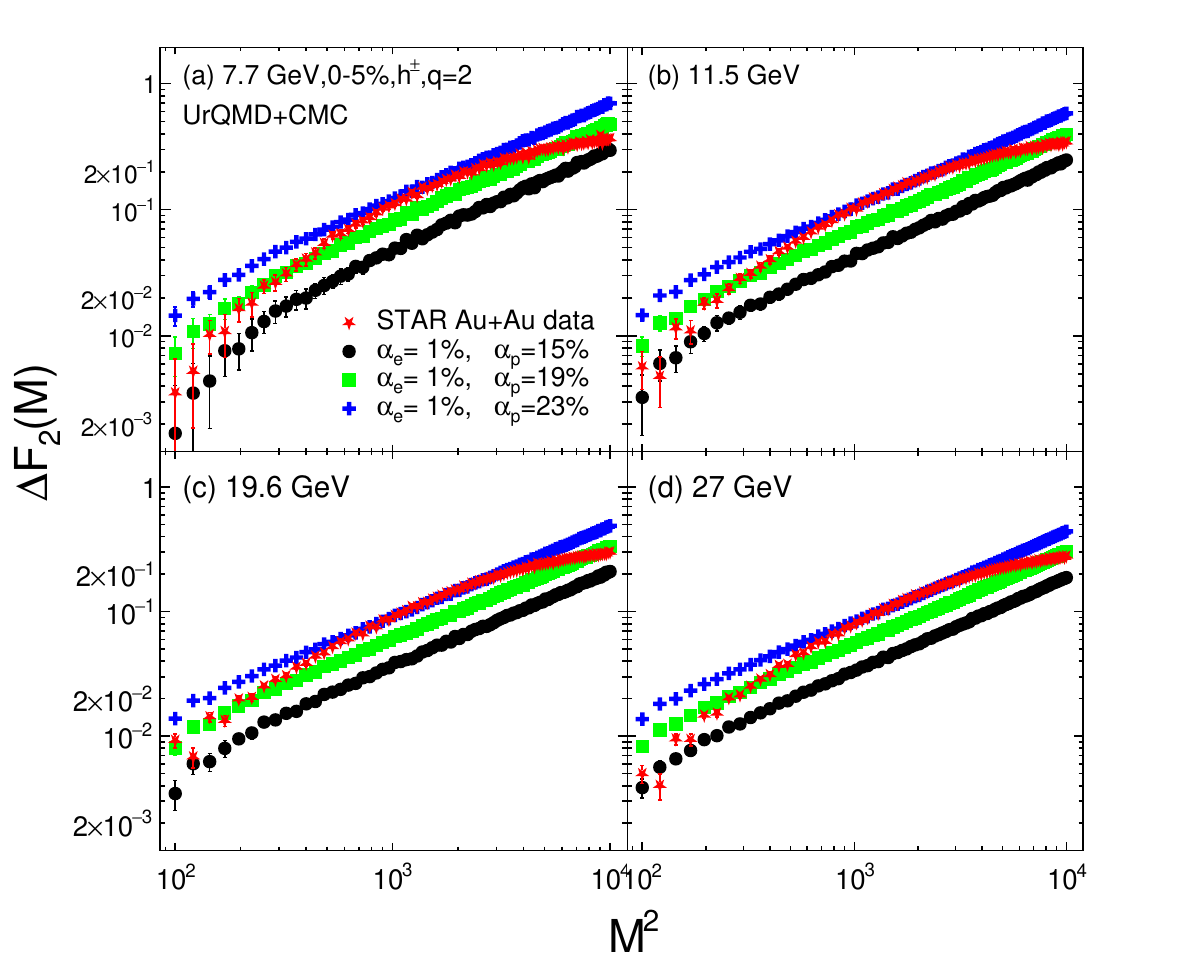}
      \caption{Comparison of the second-order correlator moment $\Delta F_2(M)$ as a function of $M^2$ for the STAR BES-I data and the hybrid UrQMD+CMC calculations with the combined event- and particle-level replacement scheme. Panels (a)--(d) show the model results for charged hadrons in 0--5\% central Au+Au collisions at $\sqrt{s_{\mathrm{NN}}}=7.7$, $11.5$, $19.6$, and $27~\mathrm{GeV}$, respectively. In all panels, the event-level fraction is fixed at $\alpha_{\mathrm{e}}=1\%$, while the particle-level fraction is varied as $\alpha_{\mathrm{p}}=15\%$ (black circles), $19\%$ (green squares), and $23\%$ (blue crosses). The red stars represent the STAR measurements~\cite{STARIntermittency}.
     }
     \label{Fig4CMCUrQMDF2Vs_STARDATA_7to27GeV}
\end{figure*}

In the particle-level replacement scheme, critical correlations are diluted within each event by the overwhelming non-critical background, whereas in the event-level replacement scheme only a very small fraction of events is assumed to contain developed critical fluctuations. As discussed in the Introduction, however, neither scheme alone provides a fully realistic physical picture. In realistic heavy-ion collisions, only a small fraction $\alpha_{\mathrm{e}}$ of events may develop a critical component, and even within those events only a fraction $\alpha_{\mathrm{p}}$ of particles may retain critical information throughout the collision evolution and remain experimentally detectable. Therefore, a realistic modeling framework should incorporate both event-level and particle-level replacement mechanisms, allowing for a more physically motivated extraction and interpretation of the intermittency signal in the STAR data.

Figure~\ref{Fig4CMCUrQMDF2Vs_STARDATA_7to27GeV} compares $\Delta F_{2}(M)$ from the STAR data with the hybrid UrQMD+CMC calculations obtained using the combined event- and particle-level replacement scheme for 0--5\% central Au+Au collisions at $\sqrt{s_{\mathrm{NN}}}=7.7$, $11.5$, $19.6$, and $27~\mathrm{GeV}$. In all panels, we fix the event-level replacement fraction at $\alpha_{\mathrm{e}}=1\%$, motivated by the previous NA49 estimate of a very small, percent-level critical-fluctuation contribution in Si+Si collisions~\cite{NA49EPJC}. The particle-level replacement fraction is varied as $\alpha_{\mathrm{p}}=15\%$, $19\%$, and $23\%$. It is found that the STAR data are generally bracketed by the calculations with $\alpha_{\mathrm{p}}=15\%$ and $23\%$, while the intermediate value $\alpha_{\mathrm{p}}\simeq 19\%$ provides the best overall description of the measured $\Delta F_{2}(M)$ over the four collision energies. Within the present combined replacement scheme, this result suggests that, under the assumption that only a small fraction of events, $\alpha_{\mathrm{e}}=1\%$, contain a critical component, approximately $19\%$ of the particles in such events may carry critical information that remains observable in the final detected state. This finding further constrains the parameter space of critical fluctuations compatible with the STAR data, indicating that even in the combined replacement scenario, the effective critical-like contribution remains limited. It should be noted that $\alpha_{\mathrm{e}}=1\%$ is a model parameter adopted in the present calculation, and its actual value in realistic heavy-ion collisions could be smaller.

For the hybrid UrQMD+CMC calculations with the event-level replacement scheme, the preferred value of $\alpha_{\mathrm{e}}$ is found to be approximately $0.04\%$ for all considered collision energies. For the particle-level replacement scheme, the preferred value of $\alpha_{\mathrm{p}}$ is approximately $1.8\%$ over the same energy range. For the combined event- and particle-level replacement scheme, the preferred value of $\alpha_{\mathrm{p}}$ is approximately $19\%$ for all considered collision energies when $\alpha_{\mathrm{e}}$ is fixed at $1\%$. Therefore, for all three schemes, the extracted critical-like fractions are very similar across the collision energies. This can be understood from two considerations. First, the STAR measurements of $\Delta F_{2}(M)$ vary only weakly with collision energy. Second, the magnitude of $\Delta F_{2}(M)$ is sensitive to variations in the critical-like fractions $\alpha_{\mathrm{p}}$ and $\alpha_{\mathrm{e}}$, which determine the fractions of the critical intermittency signal introduced into the UrQMD background at the particle and event levels, respectively. Taken together, these results suggest that the intermittency signal remains weak throughout the STAR BES-I energy range and that the effective critical-like fraction is approximately the same at all collision energies considered here. The present analysis, however, does not determine the microscopic origin of this weak intermittency signal, which may also arise from non-critical dynamical fluctuations or from non-negligible density fluctuations~\cite{JPGAntoniou,KJSunPLB2017} that develop during the evolution of heavy-ion collisions.

Recently, the STAR Collaboration measured the temperature of the hot nuclear matter created in Au+Au collisions using thermal $e^{+}e^{-}$ production~\cite{TemperatureQGP}, whose invariant-mass spectra suffer neither from strong final-state interactions nor from blue-shift effects due to rapid expansion. It was found that the temperatures extracted in the low-mass region, $T_{\mathrm{LMR}}$, from the BES-II data at $\sqrt{s_{\mathrm{NN}}}=27~\mathrm{GeV}$ and from the data at $\sqrt{s_{\mathrm{NN}}}=54.4~\mathrm{GeV}$ are in good agreement with those obtained from the BES-I measurements, as well as with the value reported by the NA60 Collaboration in In+In collisions at $\sqrt{s_{\mathrm{NN}}}=17.3~\mathrm{GeV}$. Moreover, the extracted $T_{\mathrm{LMR}}$ values remain close to one another over a broad range of collision energies, and their uncertainty ranges overlap with the QCD critical temperature ($T_{c}$). Although $T_{\mathrm{LMR}}$ and $\Delta F_{2}(M)$ probe different aspects of the heavy-ion collision system, the approximate universality of $T_{\mathrm{LMR}}$ across different $\sqrt{s_{\mathrm{NN}}}$ provides a useful qualitative context for the weak collision-energy dependence observed in the present intermittency analysis. In particular, both observations point to the absence of a pronounced, localized enhancement associated with a specific collision energy within the explored energy range.

Finally, the overall magnitude of $\Delta F_{q}(M)$ in the STAR BES-I data shows only a mild collision-energy dependence, with a weak tendency to decrease as $\sqrt{s_{\mathrm{NN}}}$ increases and with the largest values within the BES-I energy range appearing at $\sqrt{s_{\mathrm{NN}}}=7.7~\mathrm{GeV}$. This observation motivates further investigation of whether the intermittency signal exhibits any enhancement in the lower-energy STAR Fixed-Target (FXT) region, $\sqrt{s_{\mathrm{NN}}}=3$--$4.5~\mathrm{GeV}$~\cite{STARbesIIcumulant,Zhang2026SearchFT}. In this high-baryon-density region, measurements of $\Delta F_{q}(M)$ would provide a direct test of whether the weak trend observed in the BES-I data persists into the FXT energy region or whether a localized enhancement develops. A corresponding comparison with the hybrid UrQMD+CMC calculations would further allow one to examine whether the effective critical-like fractions, $\alpha_{\mathrm{e}}$ and $\alpha_{\mathrm{p}}$, remain consistent with or differ from those obtained in the BES-I $\sqrt{s_{\mathrm{NN}}}=7.7$--$27~\mathrm{GeV}$ data.

\section{Summary and Outlook}

In summary, we have performed a quantitative study to constrain the possible critical-like contribution to the intermittency signal measured in the STAR BES-I data, using an enhanced hybrid UrQMD+CMC model. Within this framework, critical-like fluctuations generated by the CMC model are embedded into a realistic non-critical UrQMD background through event-level, particle-level, and combined replacement schemes. The UrQMD+CMC calculations show that the magnitude of $\Delta F_{q}(M)$ increases systematically with the imposed critical-like fraction, thereby establishing a controlled correspondence between the measured intermittency strength and the effective critical-like fractions within the model. The STAR BES-I measurements of $\Delta F_{q}(M)$ exhibit only a weak dependence on collision energy, with a mild decreasing trend as $\sqrt{s_{\mathrm{NN}}}$ increases, while the differences among collision energies remain relatively small. This limited energy variation indicates that the intermittency strength across the BES-I energy range is confined to a narrow range and does not show a pronounced enhancement at any particular collision energy.

By performing a point-by-point comparison of $\Delta F_{2}(M)$ between the STAR BES-I data and the hybrid UrQMD+CMC calculations obtained with different replacement schemes, we constrain small and nearly energy-independent effective critical-like fractions compatible with the data. For the separate replacement schemes, the STAR measurements favor $\alpha_{\mathrm{p}}\simeq 1.8\%$ in the particle-level scheme and $\alpha_{\mathrm{e}}\simeq 0.04\%$ in the event-level scheme, while the more realistic combined scheme favors $\alpha_{\mathrm{p}}\simeq 19\%$ when $\alpha_{\mathrm{e}}=1\%$. The small and nearly energy-independent effective fractions, together with the limited variations in the STAR $\Delta F_{q}(M)$ results across BES-I energies, indicate a weak collision-energy dependence, similar to the STAR observation that the later-stage QGP temperatures $T_{\mathrm{LMR}}$ remain close to one another over a broad range of collision energies. Taken together, these findings suggest that the intermittency signal observed in the current BES-I data does not favor strong critical-point-induced fluctuations localized near a specific collision energy. Instead, the observed weak intermittency signal may predominantly reflect non-critical dynamical fluctuations in heavy-ion collisions, rather than a pronounced critical contribution.

Recent fluctuation measurements~\cite{Zhang2026SearchFT,Huang:2025wdt,STARbesIIcumulant} and elliptic-flow measurements~\cite{STAR:2025owm} from the RHIC BES-II program ($\sqrt{s_{\mathrm{NN}}}=3$--$27~\mathrm{GeV}$) have further underscored the importance of exploring the high-baryon-density region. Future intermittency measurements from the STAR Fixed-Target Program in the energy range $\sqrt{s_{\mathrm{NN}}}=3$--$4.5~\mathrm{GeV}$ would provide an important opportunity to test whether a stronger and more localized enhancement of $\Delta F_{q}(M)$ emerges, and whether the preferred particle-level and event-level critical-like fractions deviate from the BES-I values. Observation of either feature would extend the present BES-I-based study into the high-baryon-density region and thereby contribute to the ongoing effort to search for the QCD critical point.

\section* {Acknowledgments}
This work was supported by the Guangxi Natural Science Foundation under Grant No. 2025GXNSFBA069070, the Guangxi Young and Middle-aged University Teachers' (Research) Basic Competency Improvement Project under Grant No. 2025KY0533, the Fundamental Research Funds for the Central Universities under Grant No. XJ2026006701, and the National Natural Science Foundation of China under Grant Nos. 12275102 and 12135003.

\bibliography{example}	

\end{document}